\documentclass{article}
\usepackage{amsthm}
\usepackage{amsmath, amssymb, amscd}

\theoremstyle{plain}

\newtheorem{thm}{Theorem}[section]

\newtheorem{lem}{Lemma}[section]

\theoremstyle{definition}

\date{}
\numberwithin{equation}{section}
\begin{document}
\title{Hitchin Functionals and Nonspontaneous Supersymmetry Breaking}
\author{Alexander 
Golubev\footnote{e-mail address: \texttt{agolubev@nyit.edu}}\\
New York Institute of Technology, New York, NY 10023}
\maketitle
\begin{abstract}
A new mechanism of supersymmetry breaking involving a dynamical
parameter is introduced. It is independent of particle 
phenomenology and gauge groups. An explicit realization of this
mechanism takes place in Type II superstring compactifications
which admit eight supercharges (also known as generalized
$SU(3)$ structures). The resulting K\H{a}hler potentials
are expressed in terms of Hitchin functionals. Specifically, every
transversally regular generalized Calabi-Yau manifold in the moduli
space of complex or symplectic structures deforms into 
compactifications with nonspontaneously broken supersymmetry. 
Furthermore, all such deformations can be represented by elements
of the first Poisson cohomology groups of their moduli spaces, those
two groups are anti-isomorphic, and deformations of mirror pairs of 
transversally regular generalized Calabi-Yau manifolds are identified
via this anti-isomorphism. 
\end{abstract}

\section{Introduction}
The particles predicted by supersymmetric field theories failed to 
appear in experiments, so that within the accessible energy range 
there is no supersymmetry. Yet supersymmetry provides the only known
resolution of the hierarchy problem, and ensures that the Standard
Model serves as a low-energy approximation to some futuristic unified
field theory~\cite{W}. That necessitates the search for ways to make 
supersymmetry breakable (and restorable). Hitherto, this search has
produced several mechanisms of spontaneous supersymmetry breaking
~\cite{G-R, L-S, B-W, R-S}.\\
\indent
Generally speaking, it occurs only when the variation of some field under 
supersymmetry transformations yields nonzero vacuum expectation values:
\begin{equation*}
\langle \text{VAC}|\delta (\text{field})|\text{VAC} \rangle \ne 0.
\end{equation*}
\noindent
That implies a necessary condition for spontaneous supersymmetry
breaking: the generators of supersymmetry algebra must not
annihilate vacuum. However, there still remains the possibility of
all the variations having zero expectation value while the 
supersymmetry algebra generators fail to satisfy the defining
identities. Such a failure would constitute nonspontaneous 
supersymmetry breaking. And the above-mentioned identities have to
do with space-time properties of the generators. As far as we know,
the first results linking supersymmetry algebras to space-time
symmetries were published by Nahm~\cite{N}. The anti-de-Sitter
space with its $O(3,2)$ symmetries supports all conceivable
supersymmetry algebras, whereas the de-Sitter space having
$O(4,1)$ as the symmetry group has only $N=2$ supersymmetry.
Thus this Universe evolving from the anti-de-Sitter to the 
de-Sitter regime may provide a toy model of nonspontaneous
$N \ne 2$ supersymmetry breaking. It is
very instructive to expose the fatal flaw of this model. There is
no smooth direct parametric transition from $O(3,2)$ to $O(4,1)$
because $\mathfrak{o}(3,2) \ncong \mathfrak{o}(4,1)$, and for some
value of the parameter space-time symmetries collapse even
infinitesimally.\\
\indent
Therefore to make such a theory work one needs a family of locally
isomorphic Lie groups, smoothly depending on a parameter, and
differing in their facility to support supersymmetry algebras. Then
the parameter may be interpreted as the energy scale, pre- and
post-unification values separated by an interval. In the Minkowski
${\mathbb{R}}^4$ one also requires Lorentz invariance. That could
only be satisfied for families of Lie groups locally isomorphic to
the Lorentz group, and containing that group as a member. In what
follows we find one such family containing, at one extreme Spin(1, 3),
and at the other a compact Lie group $G$ which, while maintaining local
Lorentz invariance, does not 
support any supersymmetry algebras. \\
\indent
Nonspontaneous supersymmetry breaking allows for consistent pointwise
gauging in curved space-time as long as the time components of 
the curvature tensor vanish: $R^0_{\alpha \beta \xi} = R^{\alpha}_
{\beta \xi 0} = 0$. This limits its applicability, or, rather, 
underscores its purely local character.\\
\indent
Naturally, one wonders whether our mechanism is realized in 
physically relevant models. Looking at typical instances of 
spontaneous supersymmetry breaking, we notice that it relies on
field couplings in the Lagrangian. Therefore, it would seem
logical to study situations wherein some supersymmetry breaking
depends on the moduli instead of the couplings. One such situation
arises in Type II superstring backgrounds with fluxes - non-trivial
values for the NS-NS and R-R field strengths in the six-dimensional
manifolds used to compactify the theory. Introducing fluxes one 
leaves the realm of Calabi-Yau compactifications and enters into 
$SU(3)\times SU(3)$ structures. They are based on the notion of 
so-called generalized geometry due to Hitchin~\cite{Hitch}. The 
background manifolds $Y$ which replace Calabi-Yau manifolds are no
longer Ricci-flat. There are two globally defined $SU(3)$ spinors,
${\eta}^{1}$ and ${\eta}^{2}$. The holonomy group of $Y$ is,
generally speaking, not $SU(3)$ because we do not require 
Ricci-flatness. Consequently, ${\eta}^{1}$ and ${\eta}^{2}$ are not
necessarily covariantly constant. The underlying real even degree 
forms $J^{1}$, $J^{2}$ and  complex odd degree forms ${\Omega}^{1}$, 
${\Omega}^{2}$ are not closed. Their moduli
spaces ${\mathcal{M}}_J$, and ${\mathcal{M}}_{\Omega}$
are quotients of linear groups~\cite{G-L-W}, and as such
both are endowed with natural linear actions of $SO(6,6)$. The 
intrinsic torsion classes of $dJ$ and $d\Omega$~\cite{Sal} can be
used to classify all possible $SU(3)\times SU(3)$ structures. Among
those, a prominent role is played by the generalized Calabi-Yau
manifolds~\cite{Hitch}. They are precisely the backgrounds 
supporting supersymmetry~\cite{G-M-P-T}. Of generalized Calabi-Yau
manifolds, one particular subclass happens to have some remarkable
deformation properties. Namely, the transversally regular generalized
Calabi-Yau manifolds~\cite{Hitch} turn out to support supersymmetry
while allowing transversal deformations that break supersymmetry
but maintain zero expectation values of the variations of all the
fields constructed with ${\eta}^{1}$ and ${\eta}^{2}$. Thus within
a neighborhood of a transversally regular generalized Calabi-Yau
manifold in the moduli space, supersymmetry appears to be broken 
nonspontaneously. To make its breaking mechanism manifest, and to 
positively identify it as the above-proposed one, we need to study
the group-theoretical properties of transversal deformations in the 
moduli spaces. The extension of mirror symmetry inaugurated by 
Gra\~{n}a et al.~\cite{G-M-P-T} only matches pairs of generalized
Calabi-Yau manifolds (i. e. supersymmetric backgrounds).
Consequently, the transversal (supersymmetry breaking) deformations
ought to commute with mirror symmetry. This translates into a 
requirement that there be a correspondence between the transversal
deformations in the two moduli spaces. In the absense of the
$T^3$-fibration of Strominger, Yau, and Zaslow~\cite{SYZ}, the 
search for this correspondence forces us to utilize some other
structure. More specifically, we apply the notion of 
Morita equivalence of Poisson manifolds, first described by
Weinstein~\cite{We}, and Xu~\cite{Xu}, to ${\mathcal{M}}_J$ and 
${\mathcal{M}}_{\Omega}$. Using the linear $SO(6, 6)$ actions we
construct two Poisson structures ${\pi}_J$ and ${\pi}_{\Omega}$ in
such a way as to generate the leaves of their symplectic 
foliations via the linear $SO(6, 6)$ diffeomorphisms and actions
by closed two-forms ($B$-fields). Then the Casimir functions of 
${\pi}_J$ and ${\pi}_{\Omega}$ can be represented by smooth functions
of the intrinsic torsion classes. The sheaves of Casimir functions
and the first \v{C}hech cohomology groups are isomorphic via Morita
equivalence of (${\mathcal{M}}_J$, ${\pi}_J$) and 
(${\mathcal{M}}_{\Omega}$, ${\pi}_{\Omega}$) due to a theorem of 
Ginzburg and Golubev~\cite{G-G}, and from that isomorphism we are
able to deduce anti-isomorphism of the first Poisson
cohomology groups:
\begin{equation*}
H^1_{{\pi}_{\Omega}} ({\mathcal{M}}_{\Omega}) \cong
H^1_{{\pi}_J}     ({\mathcal{M}}_J).
\end{equation*}
\noindent
Given that, we assign each one-parameter family of transversal
deformations to an element of $H^1_{{\pi}} (.)$ This way they match
once the mirror pairs of transversally regular generalized 
Calabi-Yau manifolds have been chosen. As a consequence, all the
sufficiently small transversal deformations of the transversally 
regular generalized Calabi-Yau manifolds realize a representation 
of $G$, or, put in more technical terms, the representation of 
transversal deformations reduces from the most general Spin(6)
to one of its subgroups $G \subsetneq SU(4) \cong \text{Spin}(6)$.\\ 
\indent
A few words about the organization of this paper. The basic
mathematical results establishing our mechanism of supersymmetry
breaking are gathered in Section 2. The most general relativistic
setup wherein nonspontaneous supersymmetry breaking takes place 
is described in detail in Section 3. An overview of Type II 
superstring compactifications with fluxes, transversal deformations
of transversally regular generalized Calabi-Yau manifolds, their
connection with nonspontaneous supersymmetry breaking, their
representations in Poisson cohomology are the subject of Section 4.\\
\indent
Lastly, we dispense with the physical constants by setting
$ \hbar \;=\;c\;=\;1.$

\section{Mathematical Preliminaries}

The Pauli matrices are
\begin{equation}
{\sigma}_1=
\begin{bmatrix}
0&1\\
1&0
\end{bmatrix},\; {\sigma}_2=
\begin{bmatrix}
0&-i\\
i&\phantom{-}0
\end{bmatrix},\; {\sigma}_3=
\begin{bmatrix}
1&\phantom{-}0 \\
0&-1
\end{bmatrix}.
\end{equation}
\indent
The Dirac representation of $SU(2)$, denoted
$SU_{\mathcal{D}}(2)$ is generated by
\begin{equation}
J_1=\frac{1}{2}
\begin{bmatrix}
{\sigma}_1&0\\
0&{\sigma}_1
\end{bmatrix},\; J_2=\frac{1}{2}
\begin{bmatrix}
{\sigma}_2&0\\
0&{\sigma}_2
\end{bmatrix},\; J_3=\frac{1}{2}
\begin{bmatrix}
{\sigma}_3&0\\
0&{\sigma}_3
\end{bmatrix}.
\end{equation}
\noindent
There still exists the twofold covering epimorphism of Lie groups:
\begin{equation}\label{epiA}
\mathcal{A}: \;\;SU_{\mathcal{D}}(2) \longrightarrow 
\begin{bmatrix}
[SO(3)]&0\\
0&1
\end{bmatrix}.
\end{equation}
\noindent
Spin(1,3) may be viewed as a complex extension of $SU_{\mathcal{D}}(2)$:
\begin{equation}\label{Lorentzext}
\left\{ J_i = \frac{1}{2}
\begin{bmatrix}
{\sigma}_i&0\\
0&{\sigma}_i
\end{bmatrix} \right\} \mapsto
\left\{ J_i =\frac{1}{2}
\begin{bmatrix}
{\sigma}_i&0\\
0&{\sigma}_i
\end{bmatrix},\; K^{\mathbb{C}}_i=\frac{1}{2}
\begin{bmatrix}
i&\phantom{-}0 \\ 
0&-i
\end{bmatrix}
\begin{bmatrix}
{\sigma}_i&0\\
0&{\sigma}_i
\end{bmatrix}\right\}.
\end{equation}
\noindent
Fortuitously, there is a  class of mutually isomorphic almost complex 
Lie algebra extensions,
of which $\mathfrak{so}(1,3)$, generated by $\{J_i,K^{\mathbb{C}}_i\}$ 
of \eqref{Lorentzext} is a member. We are interested mainly in the
following almost complex extension:
\begin{equation}\label{Gext}
\left\{ J_i = \frac{1}{2}
\begin{bmatrix}
{\sigma}_i&0\\
0&{\sigma}_i
\end{bmatrix} \right\} \mapsto
\left\{ J_i =\frac{1}{2}
\begin{bmatrix}
{\sigma}_i&0\\
0&{\sigma}_i
\end{bmatrix},\; K_i=\frac{1}{2}
\begin{bmatrix}
\phantom{-}0&1 \\ 
-1&0
\end{bmatrix}
\begin{bmatrix}
{\sigma}_i&0\\
0&{\sigma}_i
\end{bmatrix}\right\}.
\end{equation}
\noindent
Its relevant properties are summarized in
\begin{thm}\label{Gcompactness}
There exists a unique compact semisimple Lie group $G \subset SU(4)$,
whose Lie algebra $ \mathfrak{g}\cong \mathfrak{so}(1,3)$ is generated
by \eqref{Gext}.
\end{thm}
\begin{proof}
Every almost complex extension corresponds (up to a nonzero factor)
to a matrix
\begin{equation*}
\begin{bmatrix}
a&b\\
c&d
\end{bmatrix}\in U(2),\;\;
\begin{bmatrix}
a&b\\
c&d
\end{bmatrix}
\begin{bmatrix}
a&b\\
c&d
\end{bmatrix}=
\begin{bmatrix}
-1&\phantom{-}0 \\ 
\phantom{-}0&-1
\end{bmatrix}.
\end{equation*}
\noindent
$ \mathfrak{g}\cong \mathfrak{so}(1,3)$ implies $ad -bc=1$.
Therefore
\begin{equation*}
\Re a = \Re d=0,\; \Im c = \Im b, \; \Re c =- \Re b.
\end{equation*}
\noindent
This allows us to write the most general almost complex
extension as
\begin{equation*}
J_i \mapsto \left(w
\begin{bmatrix}
i&\phantom{-}0 \\ 
0&-i
\end{bmatrix} + u
\begin{bmatrix}
0&i \\ 
i&0
\end{bmatrix} + v
\begin{bmatrix}
\phantom{-}0&1\\
-1&0
\end{bmatrix}\right)J_i,\; w^2 + u^2 + v^2 =1.
\end{equation*}
\indent
To ensure compactness, we must have $\exp i{\kappa}^aK_a$ bounded. 
Whence $w=0$, $u=0$ is the only choice. And this is \eqref{Gext}.\\
\indent
According to Helgason (\cite{He}, Chapter~II, \S 2, Theorem~2.1), 
there exists a Lie group $G$, whose Lie algebra is generated by 
$\{J_i, K_i\}$ of \eqref{Gext}. Its elements are all of the form
$\exp i({\theta}^bJ_b + {\kappa}^a K_a)$, which means $G$ is a 
Lie subgroup of $SU(4)$.
Now $G$ has to be closed in the standard matrix topology of $SU(4)$.
That is based on a fundamental
result of Mostow~\cite{M}: any semisimple Lie subgroup $H$ of a compact 
Lie group $C$ is closed in the relative topology of $C$. In our case,
$SU(4)$ is compact, $\mathfrak{g}$ is semisimple.
\end{proof}
\indent
In the sequel we will work with the homogeneous space 
$G/SU_{\mathcal{D}}(2)$.
\begin{lem}\label{trivialhomotopy}
\begin{equation*}
{\pi}_1(G/SU_{\mathcal{D}}(2))= 0.
\end{equation*}
\end{lem}
\begin{proof}
For all Lie groups ${\pi}_2 (.) = 0$~\cite{B}; for 
$SU_{\mathcal{D}}(2)$, 
${\pi}_0(SU_{\mathcal{D}}(2))= 0$ by connectedness. Also, 
$SU_{\mathcal{D}}(2)$ is a
closed subgroup of $SU(4)$ in the ordinary matrix topology.
We therefore have the following exact homotopy sequence~\cite{B}:
\begin{equation*}
\begin{split}
0 \rightarrow {\pi}_2 (SU(4)/SU_{\mathcal{D}}(2))& 
\rightarrow {\pi}_1 (SU_{\mathcal{D}}(2))\\
&\rightarrow {\pi}_1 (SU(4)) \rightarrow {\pi}_1 
(SU(4)/SU_{\mathcal{D}}(2))\rightarrow 0.
\end{split}
\end{equation*}
\noindent
${\pi}_1 (SU(4))= 0$~\cite{B} whence 
\begin{equation*}
{\pi}_1 (SU(4)/SU_{\mathcal{D}}(2)) \cong {\pi}_1 
(SU_{\mathcal{D}}(2)) = {\pi}_1({\mathbb{S}}^3) =0.
\end{equation*}
\noindent
Now homotopy is functorial. The embedding
$\xi : G/SU_{\mathcal{D}}(2) \hookrightarrow 
SU(4)/SU_{\mathcal{D}}(2)$ 
induces the monomorphism of fundamental groups
\begin{equation*}
{\xi}_{\pi *} : {\pi}_1 (G/SU_{\mathcal{D}}(2)) \rightarrow
{\pi}_1 (SU(4)/SU_{\mathcal{D}}(2)). \qedhere
\end{equation*}
\end{proof}
\begin{thm}
\begin{equation*}\label{s3theorem}
 G/SU_{\mathcal{D}}(2) \cong {\mathbb{S}}^3.
\end{equation*}
\end{thm}
\begin{proof}
\noindent
$\mathfrak{g}$ 
decomposes as a vector space into two three-dimensional
subspaces,
\begin{equation*}
\mathfrak{g} = \mathfrak{j} \oplus \mathfrak{k},
\end{equation*}
\noindent
Based on this decomposition, there is an involutive automorphism 
\begin{equation*}
\vartheta : \mathfrak{g} \;
\longrightarrow \;
\mathfrak{g} 
\end{equation*}
\noindent
defined by
\begin{equation*}
\vartheta (J + K) = J - K,\quad \forall J \in \mathfrak{j},
\quad \forall K \in \mathfrak{k}.
\end{equation*}
\noindent
$\mathfrak{j}$ is the set of fixed points of $\vartheta$. It is 
unique (\cite{He}, Chapter IV, \S 3, Proposition 3.5). The pair 
$(\mathfrak{g} \;,\; \vartheta)$ is an orthogonal symmetric Lie algebra
(\cite{He}, Chapter~IV, \S 3). There is a Riemannian symmetric
pair $(G,\; SU_{\mathcal{D}}(2))$ associated with $(\mathfrak{g}, \; 
\vartheta)$ so that the quotient
$G/SU_{\mathcal{D}} (2)$ is a complete locally symmetric Riemannian space.
Furthermore, its curvature corresponding to any $ G$-invariant
Riemannian structure is given by (\cite{He}, Chapter IV, \S 4, 
Theorem 4.2):
\begin{equation*}
R(K_{i_1}, K_{i_2})K_{i_3}= -[[K_{i_1}, K_{i_2}], K_{i_3}]
\quad \forall K_{i_1}, K_{i_2}, K_{i_3} \in \mathfrak{k}.
\end{equation*}
\noindent
Computing the sectional curvature we see that $R^{\text{sect}} 
\equiv 1$. Now a pedestrian version of the Sphere 
theorem~\cite{C-G} asseverates that a complete simply connected 
Riemannian manifold with $R^{\text{sect}} \equiv 1$ is isometric 
to a sphere of appropriate dimension. In our case the topological
condition is satisfied in view of Lemma~\ref{trivialhomotopy}.
\end{proof}
\indent
Consider the natural inclusions of Lie groups
\begin{equation}
\iota : G \hookrightarrow GL(4, \mathbb{C}), \quad
\iota : \text{Spin}(1, 3) \hookrightarrow GL(4, \mathbb{C}).
\end{equation}
\noindent
Their images inside $GL(4, \mathbb{C})$ intersect:
\begin{equation}\label{properspin}
\iota (G) \cap \iota (\text{Spin}(1, 3))=SU_{\mathcal{D}}(2).
\end{equation} 
\noindent
Because of~\eqref{properspin}, the set 
\begin{equation}
{\text{Ad}}_{\iota (G)}(\iota (\text{Spin}(1, 3)))=
\coprod_{U \in G} U\text{Spin}(1, 3)U^H, 
\end{equation}
the disjoint union of conjugates of $\text{Spin}(1, 3))$,
has the same cardinality as the set of 
all boosts in $G$. Similarly, there is the natural inclusion
\begin{equation}
\iota :\quad SO(4) \hookrightarrow GL(4, \mathbb{R}).
\end{equation}
\noindent
The set ${\text{Ad}}_{\iota (SO(4))}(\iota (\text{SO}(1, 3)^e))$ is
homeomorphic to $SO(4)/SO(3) \cong {\mathbb{S}}^3$. Combining this with
Theorem~\ref{s3theorem} we arrive at:
\begin{equation}\label{wpdiff}
\begin{CD}
\text{Ad}_{\iota (G)}(\iota (\text{Spin}(1,3))) @=
G/SU_{\mathcal{D}} (2) @>{\cong}>> {\mathbb{S}}^3\\ 
@.             @V{\wp}VV                 @|\\
\text{Ad}_{\iota (SO(4))}(\iota (\text{SO}(1,3)^e))
@= SO(4)/SO(3) @>{\cong}>>  {\mathbb{S}}^3
\end{CD}
\end{equation}
\noindent
The double horizontal lines indicate set-theoretic bijective 
correspondences, the upper $\cong$ is an isometry, the lower one 
is a diffeomorphism. Furthermore, the diagram~\eqref{wpdiff} 
commutes and \textit{de facto} defines the diffeomorphism $\wp$.
This diffeomorpism is utilized in the sequel to effect an
action of $G$.

\section{Equivariant Momentum Operators}

\indent
$G$ does not act on the Minkowski ${\mathbb{R}}^4$ by isometries.
We have
\begin{equation}
\begin{cases}
G \times {\mathbb{R}}^4 \longrightarrow {\mathbb{R}}^4;\\
(\exp i({\theta}^b J_b +{\kappa}^a K_a), x^{\mu}) \mapsto
x'^{\mu}= {\mathcal{A}(\exp i{\theta}^b J_b)}^{\mu}_{\lambda}
{\wp(\exp i{\kappa}^a K_a)}^{\lambda}_{\eta} x^{\eta}.
\end{cases}
\end{equation}
\noindent
In fact, the metric becomes frame-dependent:
\begin{equation*}
\wp (\exp i{\kappa}^a (\alpha) K_a) g \wp (\exp(- i{\kappa}^a (\alpha) K_a))
\ne g, \; \alpha \in [0, 2\pi],\; \alpha \ne \{0, 2\pi \}; 
\end{equation*}
$\alpha$ being the group parameter here. Yet physical
quantities must remain frame-independent. Therefore,
instead of the standard quantum field theory substitution
\begin{equation}\label{standardQFT}
P_{\mu} \longrightarrow i{\partial}_{\mu},
\end{equation}
\noindent
we employ the rule
\begin{equation}\label{nabla}
P_{\mu} \longrightarrow i{\nabla}_{\mu}({\alpha}) 
\overset{\text{def}}{=} i({\varepsilon}^{\nu}_{\mu}(\alpha)
{\partial}_{\nu} + i{\kappa}_{\mu}^a (\alpha) K_a),
\end{equation}
\noindent
the exact form of ${\varepsilon}^{\nu}_{\mu}(\alpha)$ and
${\kappa}_{\mu}^a (\alpha)$ to be determined. $K_a$'s are
in keeping with the (1/2, 1/2) representation of $P_{\mu}$'s.
This construction is an equivariant incarnation of the free
spin structure due to Plymen and Westbury~\cite{P-W}. 
Briefly, let $M$ be a 4-dimensional 
smooth manifold with all the obstructions to the existence of 
a Lorentzian metric vanishing (for instance, a parallelilazable 
$M$ would do). Let 
\begin{equation*}
\Lambda : \;\; \textnormal{Spin} (1, 3) \rightarrow SO(1, 3)^e
\end{equation*}
be the twofold covering epimorphism of Lie groups.
A free spin structure on $M$ consists of a principal bundle 
$\zeta : \Sigma \rightarrow M$ with
structure group $\textnormal{Spin} (1, 3)$ and a bundle map
$\widetilde{\Lambda}: \Sigma \rightarrow \mathcal{F} M$ into the bundle
of linear frames for $TM$, such that
\begin{equation*}
\widetilde{\Lambda} \circ {\widetilde{R}}_S = 
{\widetilde{R}}'_{\iota \circ \Lambda (S)} \circ \widetilde{\Lambda}
\;\;\; \forall S \in \textnormal{Spin} (1, 3),
\end{equation*}
\begin{equation*}
{\zeta}' \circ \widetilde{\Lambda} = \zeta,
\end{equation*}
${\widetilde{R}}$ and ${\widetilde{R}}'$ being the canonical right actions
on $\Sigma$ and $\mathcal{F} M$ respectively, $\iota : SO(1,3)^e \rightarrow
GL(4, \mathbb{R})$ the natural inclusion of Lie groups, and ${\pi}' :
\mathcal{F} M \rightarrow M$ the canonical projection. The map 
$\widetilde{\Lambda}$ is called a spin-frame on $\textnormal{Spin}(1, 3)$.
This definition of a spin structure induces metrics on $\Sigma$. Indeed,
given a spin-frame $\widetilde{\Lambda}: \Sigma \rightarrow \mathcal{F} M$,
a dynamic metric $g_{\widetilde{\Lambda}}$ is defined to
be the metric that ensures orthonormality of all frames in 
$\widetilde{\Lambda}(\Sigma) \subset \mathcal{F} M$. It should be
emphasized that within the Plymen and Westbury's formalism the 
metrics are built \textit{a posteriori},
after a spin-frame has been set by the field equations. In our
formalism the metrics are obtained via the $G$-action, and the
set of all allowable metrics is $\text{Ad}_{\iota (SO(4))}
(\iota (\text{SO}(1,3)^e))$.\\
\indent
${\nabla}_{\mu}({\alpha})$ qualifies as a $G$-connection on the 
principal $G$-bundle over the physical space-time. Furthermore,
we impose an additional condition on~\eqref{nabla} to ensure
validity of the relativistic impulse-energy identity:
\begin{equation}\label{coinvariance}
{P}^{\mu}(\alpha){P}_{\mu}(\alpha)= g^{\nu \lambda}({\alpha})
{\nabla}_{\nu}({\alpha}){\nabla}_{\lambda}({\alpha})
\overset{\text{def}}{=} g^{\nu \lambda}(0)
{\partial}_{\nu}{\partial}_{\lambda}= P^{\mu}(0)P_{\mu}(0).
\end{equation}
\noindent
This translates to some algebraic relations among ${\kappa}^a_{\mu}$'s
and ${\varepsilon}^{\nu}_{\mu}$'s. However, we still need to make
the $G$-transformation law of~\eqref{nabla} more explicit. First,
these operators are natural spinors in the sense that $SU_{\mathcal{D}}(2)$
acts linearly:
\begin{alignat}{2}\label{su2action}
U{\gamma}^{\mu}{\nabla}_{\mu}U^H &=U{\gamma}^{\mu}U^H
{\varepsilon}^{\nu}_{\mu}{\partial}_{\nu} + i{\kappa}_{{\mu}}^a
U{\gamma}^{\mu}U^HUK_aU^H& &\quad\\
&=M^{\mu}_{\eta}{\gamma}^{\eta}{\varepsilon}^{\nu}_{\mu}{\partial}_{\nu}
+M^{\mu}_{\eta}{\gamma}^{\eta}i{\kappa}_{{\mu}}^a r^n_a K_n & &
\quad \text{by}\;[\mathfrak{j}, \mathfrak{k}] = \mathfrak{k}.\notag
\end{alignat}
\noindent
Here $M^{\mu}_{\eta}$'s realize an $SO(3)$ transformation 
$(U \in SU_{\mathcal{D}}(2))$, which is
at its most transparent if ${\gamma}^{0}$ is diagonal. 
As for $r_a^n$'s, they determine how the potentials behave:
\begin{equation}
{\Tilde{\kappa}}^a_{\mu}={\kappa}_{{\mu}}^1 r^a_1 +
{\kappa}_{{\mu}}^2 r^a_2 + {\kappa}_{{\mu}}^3 r^a_3, \quad {\text{and}}
\end{equation}
\begin{equation}
{|r^a_1|}^2 +{|r^a_2|}^2 +{|r^a_3|}^2=1,\quad a=\{1,2,3\}.
\end{equation} 
\noindent
To see how they are boosted, we treat a prototypical case - that of a
boost in the $x^{3}$-direction. Specifically,
\begin{align}
{\nabla}_{0}&={\varepsilon}^{0}_{0}(\alpha){\partial}_0 + 
{\varepsilon}^{3}_{0}(\alpha){\partial}_3 + i{\kappa}_0(\alpha)K_3,\\
{\nabla}_{3}&={\varepsilon}^{0}_{3}(\alpha){\partial}_0 + 
{\varepsilon}^{3}_{3}(\alpha){\partial}_3 + i{\kappa}_3(\alpha)K_3,\\
{\nabla}_{1} &= {\partial}_1,\\
{\nabla}_{2} &= {\partial}_2.
\end{align}
\noindent
We look for solutions of 
\begin{equation}\label{Dirac1}
(i {\gamma}^{\mu}{\nabla}_{\mu} - m)\Psi = 0,
\end{equation}
\noindent
modelled on the free spinors
\begin{equation}
\Psi(\alpha)= s(\alpha)e^{-i(p_0x^0 +p_3x^3)},
\end{equation}
subject to the relativistic impulse condition ${p_0}^2-{p_3}^2=m^2$.
In the standard representation
\begin{equation}
{\gamma}^0 =
\begin{bmatrix}
1&\phantom{-}0\\
0&-1
\end{bmatrix},\quad {\gamma}^{i} =
\begin{bmatrix}
0&-{\sigma}_i\\
{\sigma}_i&\phantom{-}0
\end{bmatrix},
\end{equation}
\noindent
the equation \eqref{Dirac1} yields the following matrix:
\begin{equation}
\begin{bmatrix}
{\varepsilon}_{0}(\alpha)-m(\alpha)
&0&-{\varepsilon}_{3}(\alpha)-{\kappa}_{0}(\alpha)&0\\
0&{\varepsilon}_{0}(\alpha)-m(\alpha)
&0&{\varepsilon}_{3}(\alpha)+{\kappa}_{0}(\alpha)\\
{\varepsilon}_{3}(\alpha)-{\kappa}_{0}(\alpha)
&0&-{\varepsilon}_{0}(\alpha)-m(\alpha)&0\\
0&-{\varepsilon}_{3}(\alpha)+{\kappa}_{0}(\alpha)
& 0&-{\varepsilon}_{0}(\alpha)-m (\alpha) 
\end{bmatrix}, 
\end{equation}
where the entries are
\begin{align}
{\varepsilon}_{0}(\alpha)&={\varepsilon}^{0}_{0}(\alpha)p_0 + 
{\varepsilon}^{3}_{0}(\alpha)p_3,\\
{\varepsilon}_{3}(\alpha)&={\varepsilon}^{0}_{3}(\alpha)p_0 + 
{\varepsilon}^{3}_{3}(\alpha)p_3,\\
m(\alpha)&= m +{\kappa}_{3} (\alpha).  
\end{align}
\noindent
Its rank has to be 2 for all values of $\alpha$, thus 
constraining ${\kappa}_0(\alpha)$ and ${\kappa}_3(\alpha)$:
\begin{equation}
{{\varepsilon}_{0}}^2(\alpha) -  {{\varepsilon}_{3}}^2(\alpha)=
(m +{\kappa}_{3}(\alpha))^2 -{{\kappa}_0}^2(\alpha).
\end{equation}
\noindent
Evidently ${\kappa}_{\mu}^a(\alpha)$'s are not identically zero. At the
same time, ${\kappa}_{\mu}^a(0) =0, \;\forall \mu = \{0,1,2,3\}$.
Hence, a boost entails a nonlinear change in the potentials.\\
\indent
Finally, we are in a position to deal with supersymmetry algebras. For
the reminder of this section, the impulse operators and all other
quantities expressly depend on the parameters introduced in the proof
of Theorem~\ref{Gcompactness}. For convenience, we bundle them into
one complex parameter $z$ via stereographic projection, so that
$K_a(0)= K_a^{\mathbb{C}}$, $K_a(1)=K_a$, ${\varepsilon}^{\nu}_{\mu}
(\alpha, 0)={\delta}^{\nu}_{\mu}$, ${\kappa}_{\mu}^a (\alpha, 0)=0$.  
Should there exist such algebras,
$ Q_m(z), {\bar{Q}}_m(z)$ would generate them. But they 
realize a linear representation of the (respective) symmetry group, and
we arrive at an equality impossible for some $z \in [0,1]$:
\begin{equation}
\{ Q_m(z), {\bar{Q}}_m(z) \} = 
-2i{\gamma}^{\mu}({\varepsilon}^{\nu}_{\mu}(\alpha, z)
{\partial}_{\nu} + i{\kappa}_{\mu}^a (\alpha, z) K_a(z)).
\end{equation}
\noindent
The right-hand side transforms nonlinearly because of 
${\kappa}_{\mu}^a (\alpha,1)$, whereby proving that there are
no $ Q_m(1), {\bar{Q}}_m(1)$. Adding central charges $Z_m$,
$Z^*_m$ on the right-hand side would not remedy the situation 
because these charges commute with the symmetry group generators.\\
\indent
The next question we address is that of gauging nonspontaneous
supersymmetry breaking in curved space-time. Pseudo-Riemannian
geometry has two ways to account for variability of the metric.
One way is to introduce the Levi-Civita connection and the
curvature tensor. The alternative is the Cartan's method of 
equivalence~\cite{Gard}. The gist of the latter consists in 
specifying the subgroups of local diffeomorphisms that preserve
the geometry. Thus the most general local diffeomorphism induces
an automorphism of the tangent bundle $a^*(x) \in SO(1, 3)^e 
\times {\mathbb{R}}^4$. In our
case we further constrain local diffeomorphisms by insisting that 
$a^*$ preserve the $G$-connection. The inherent
gauge freedom of the connection allows it to transform
equivariantly. Hence pointwise $a^* \in (SO(1, 3)^e \cap \rho(G))$, 
viewed as a set of abstract automorphisms, $\rho(G)$ being some
linear representation of $G$. According to \eqref{su2action},
\begin{equation}
\rho|_{SU_{\mathcal{D}}(2)} =  \mathcal{A} \;\;\; \text{of \eqref{epiA}},
\end{equation}
\noindent
and the subgroup of spatial rotations - $SO(3)$ - does afford
equivariance. In the language of curvature tensor coefficients
these bundle automorphisms correspond to $R^0_{\alpha \beta \xi} 
= R^{\alpha}_{\beta \xi 0} = 0$.
To name a few examples, the expanding Friedman universe can have
nonspontaneously broken supersymmetry. By contrast, in the 
mixmaster universe supersymmetry is either intact or broken
spontaneously. Finally, in the most realistic unevenly expanding
universe with deviations from the spherical symmetry, there are
causally disconnected patches of supersymmetry in an otherwise
asupersymmetric space-time, because, on general grounds, it is 
hard to see how supersymmetry can be broken spontaneously in
all those patches.

\section{Not So Super Strings}

\subsection{Type II Compactifications}
Our goal now is to place nonspontaneous supersymmetry breaking in 
the context of Type II superstring theories. They play out on the
space-time background of ten dimensions. The underlying manifolds
have the pseudo-Riemannian metric of signature (1,9).
We specialize $M^{1,9}$ to have a fixed tangent bundle splitting
\begin{equation}\label{split}
TM^{1,9} = T{\mathbb{R}}^{1,3} \oplus TY,
\end{equation}
augmented with the requirement that there be a local smooth fibration
\begin{equation}\label{fibration}
f_O: M^{1,9} \searrow O \subset {\mathbb{R}}^{1,3}
\end{equation}
over the physical space-time, each fiber $f^{-1}_O(x)$ being a compact 
6-dimensional manifold. The splitting~\eqref{split} implies a 
decomposition of the Lorentz group Spin(1,9) $\supset$ Spin(1,3)
$\times$ Spin(6) and an associated decomposition of the spinor
representation $\textbf{16} \in$ Spin(1,9) according to
$\textbf{16} \rightarrow (\textbf{2}, \textbf{4}) \oplus
(\bar{\textbf{2}}, \bar{\textbf{4}})$. $TY$ is a $SO(6)$ vector
bundle which admits a pair of distinct $SU(3) \subset$ Spin(6)
structures
${\eta}^1$, ${\eta}^2$. Now ${\eta}^{1,2} =({\eta}^{1,2}_{+},
{\eta}^{1,2}_{-})$, where the '+' and '-' signs indicate the
6-dimensional chirality. Given the set of 6-dimensional
$\gamma$-matrices ${\mathring{\gamma}}^m,\;m=\{1,2,...,6\}$, 
we have the following identities:
${\eta}^{1,2}_{-}=({\eta}^{1,2}_{+})^c$, 
where ${\eta}^c= D{\eta}^*$, $D$ being the intertwiner: 
${\mathring{\gamma}}^{m*} =D^{-1}{\mathring{\gamma}}^m D$. Also,
${\bar{\eta}}={\eta}^{\dagger}B$, where ${\mathring{\gamma}}^
{m\dagger} =B {\mathring{\gamma}}^m B^{-1}$. In the sequel we'll
need the chirality operator, denoted ${\mathring{\gamma}}$. It 
assigns signs via ${\mathring{\gamma}}{\eta}^{1,2}_{\pm}=
\pm {\eta}^{1,2}_{\pm}$.\\
\indent
The Clifford algebra of $Y$ is a deformation of 
${\Lambda}^*T^*Y \otimes \mathbb{C}$.
Utilizing the Fierz map we identify:
\begin{equation}
{\eta}_{+}^{1,2} \otimes {\bar{\eta}}^{1,2}_{+} = e^{iJ^{1,2}},
\end{equation}
\begin{equation}
{\eta}_{+}^{1,2} \otimes {\bar{\eta}}^{1,2}_{-} = {\Omega}^{1,2}.
\end{equation}
Here $J$ is a real even-degree form, $\Omega$ is a
complex odd-degree form, and the spinors are normalized so that
${\eta}_{\pm}^{1,2} {\bar{\eta}}^{1,2}_{\pm}= \frac{1}{2}$. \\
\indent
In the presence of fluxes these spinors are no longer
covariantly constant. Therefore the attendant differential
forms are not closed. Instead we have~\cite{Sal}:
\begin{equation}\label{Itorsion1}
dJ = \frac{3}{4}i (W_1\bar{\Omega}-{\bar{W}}_1 \Omega) +
W_4 \wedge J + W_3,
\end{equation}
\begin{equation}
d\Omega = W_1 J\wedge J + W_2 \wedge J + {\bar{W}}_5 \wedge\Omega,
\end{equation}
\begin{equation}\label{Itorsion2}
W_3 \wedge J = W_3 \wedge \Omega = W_2 \wedge J \wedge J=0.
\end{equation}
\noindent
In the last three formulas $W_1$ is a zero-form, $W_4$, $W_5$
are one-forms, $W_2$ is a two-form, and $W_3$ is a three-form,
and we omit the superscripts. In addition, we have
\begin{equation}
J^{1,2} \wedge J^{1,2}\wedge J^{1,2} = \frac{3}{4} i\,
{\Omega}^{1,2}  \wedge {\bar{\Omega}}^{1,2}, \;\;\;\; 
J^{1,2} \wedge {\Omega}^{1,2} =0.
\end{equation}
\indent
When ${\eta}^1={\eta}^2$, $dJ^{1,2}=d{\Omega}^{1,2}=0$, $J$
is a symplectic form, $\Omega$ determines the complex
structure, and their combination defines a Calabi-Yau manifold.\\
\indent  
Geometrically, ${\eta}^1$, ${\eta}^2$ provide  two different 
decompositions of the complexified tangent bundle:
\begin{equation}
T_{\mathbb{C}}Y = TY \otimes \mathbb{C} = E^1 \oplus {\bar{E}}^1 =
E^2 \oplus {\bar{E}}^2.
\end{equation} 
\noindent
$E^1$ and $E^2$ specify two different embeddings $SU(3) 
\overset{{\tau}_{1,2}}{\hookrightarrow}$ Spin(6). Now the existence
of the smooth local fibration signifies that $\frac{\partial}
{\partial x^{\mu}}{\tau}_{1,2} \ne 0$ and the transformation law
of $SU(3) \times SU(3)$ structures takes the entire Spin(6)
$\times$ Spin(6) group. In contrast, a global fibration
$M^{1,9} = {\mathbb{R}}^{1,3} \times Y$ would have kept ${\tau}_{1,2}$
constant throughout ${\mathbb{R}}^{1,3}$.\\
\indent
Alternatively, recall that the original Type II theory is formulated
on a supermanifold $M^{1,9|16+16}$ of bosonic dimension (1,9) with a
manifest local Spin(1,9) symmetry and with the Grassmann variables
transforming as a pair of 16-dimensional spinor representations. The
requirement that we have an $SU(3) \times SU(3)$ structure means 
there is a sub-supermanifold $N^{1,9|4+4} \subset M^{1,9|16+16}$,
still of bosonic dimension (1,9), but now with only eight Grassmann
variables transforming as spinors of Spin(1,3) and singlets of one
or the other $SU(3)$. It is natural to reformulate the ten-dimensional
theory in this sub-supermanifold.\\

\subsection{Generalized Structures}
The generalized structures of Hitchin~\cite{Hitch} are associated with
a subgroup of $SO(n,n)$. This is the structure group of the bundle
$TY \oplus T^*Y$, which always exists. It is a rank $2n$ vector bundle
with a preferred orientation and an inner product/metric of signature
$(n,n)$:
\begin{equation}
(X+ \xi, X + \xi) =- i_X \xi =-X_l {\xi}^l, \;\;\; l=\{1, ..., n\}.
\end{equation}
In conjunction with this metric, there always is a spin structure in 
view of standard properties of the Stiefel-Whitney classes $w_{\sharp}
\in H^{\sharp}(Y,{\mathbb{Z}}_2)$:
\begin{equation*}
w_2(TY \oplus T^*Y) = w_2(TY) +w_1(TY)\cup w_1(T^*Y) +w_2(T^*Y)=0
\end{equation*}
\noindent
(using $w_{\sharp}(TY) =w_{\sharp}(T^*Y)$). The corresponding spinors,
sections of the spin bundles $S^{\pm}$ of Spin($n,n$) are isomorphic
to even or odd forms on $Y$. However, the isomorphism is not canonical.
Rather, it is parameterized by sections of $GL(n)/SL(n)$, that is by
a scalar field (dilatino) $e^{-\phi}, \;\phi \in C^{\infty}(Y)$:
\begin{equation}
S^{\pm} = {\Lambda}^{\text{even/odd}}T^*Y \otimes \sqrt{{\Lambda}^n TY},
\;\;\; \phi (y): \sqrt{{\Lambda}^n T_{y}Y} \longrightarrow \mathbb{R}. 
\end{equation}
\noindent
As a matter of notation, $\Phi \in S^{\pm}$ is designated
$\Phi = \theta \otimes e^{-\phi}$.\\
\indent
In our case the group is $SO(6,6)$. We write it as $SO(V \oplus V^*)$,
$\dim V =6$. Its Lie algebra has the following decomposition:
\begin{equation}
\mathfrak{so} (V \oplus V^*) =
V \otimes V^* \oplus {\Lambda}^2 V^* \oplus {\Lambda}^2 V.
\end{equation}
\noindent
$V \oplus V^*$ acts on the Clifford algebra via
\begin{equation}\label{Cliffaction}
(X + \xi)\cdot \Phi = i_X \Phi + \xi \wedge \Phi.
\end{equation}
\indent
There is an invariant skew-symmetric bilinear form on $S^{\pm}$ known
in the physics literature as the Mukai pairing:
\begin{equation}
< \Phi, \Psi> = \sum_{m} (-1)^m {\Phi}_{2m} \wedge {\Psi}_{6-2m}
\in {\Lambda}^6 V^* \otimes (\sqrt{{\Lambda}^6 V})^2 = \mathbb{R}
\end{equation}
for even spinors, and
\begin{equation}
< \Phi, \Psi> = \sum_{m} (-1)^m {\Phi}_{2m+1} \wedge {\Psi}_{6-2m-1}
\in {\Lambda}^6 V^* \otimes (\sqrt{{\Lambda}^6 V})^2 = \mathbb{R}
\end{equation}
for odd ones. $<\cdot, \cdot>$ is symplectic and allows us to
identify the linear structure of $S^{\pm}$ with that of $TS^{\pm}$.\\
\indent
Given $\Phi \in S^{\pm}$, consider its annihilator, the vector
space
\begin{equation}
V_{\Phi} =\{ X + \xi \in V\oplus V^*|(X +\xi) \cdot \Phi =0 \}.
\end{equation}
\noindent
A spinor $\Phi$ for which $\dim V_{\Phi} =6$ is called a pure
spinor. Any two pure spinors are related by an element of 
Spin(6,6).\\
\indent
The generalized Calabi-Yau manifolds are characterized as pairs
$(Y, \Phi)$, $\Phi$ being a pure complex spinor satisfying
$< \Phi,\bar{\Phi}> \ne 0$ at each point and such that the
corresponding differential form, a section of one of the
complexified exterior bundles ${\Lambda}^{\textnormal{even/odd}}
T^* Y \otimes \mathbb{C}$, is closed. The class of generalized
Calabi-Yau manifolds trivially includes classical Calabi-Yau
as well as symplectic manifolds.\\
\indent
For the time being, we work with the semisimple group 
Spin(12, $\mathbb{C}$) instead of Spin(6,6) and a complex
six-dimensional vector space $V$. On each of the 32-
dimensional spin spaces $S^{\pm}$, the actions of both
groups are symplectic. Consequently, there is the moment map
$\mu : S^{\pm} \longrightarrow {\mathfrak{so}}(12, \mathbb{C})^{*}$:
\begin{equation}
\mu(\Phi)(X) = \frac{1}{2}<\rho (X)\Phi, \Phi>, 
\end{equation}
where $\Phi \in S^{\pm},\;\;\rho :{\mathfrak{so}}(12, \mathbb{C})
\rightarrow \text{End}S^{\pm}$ denotes the representations of 
${\mathfrak{so}}(12, \mathbb{C})$, and $X \in 
{\mathfrak{so}}(12, \mathbb{C})$. Using the nondegeneracy of 
the Killing bilinear form we identify ${\mathfrak{so}}(12, 
\mathbb{C})$ with its dual, so $\mu(\Phi)$ takes values in the
Lie algebra. Now we define an invariant quartic function on 
$S^{\pm}$ via
\begin{equation}
q(\Phi) \overset{\text{def}}{=} \text{tr} \; \mu(\Phi)^2.
\end{equation}
\noindent
For $\Phi \in S^{\pm}$, $q(\Phi) \ne 0$ if and only if 
$\Phi = \alpha + \beta$, where $\alpha$ and $\beta$ are
pure spinors such that $<\alpha, \beta> \ne 0$. We are
interested in the open sets
\begin{equation}
S^{+}_q = \{ \Phi \in S^{+}|q(\Phi) <0\},\;\;\;
S^{-}_q = \{ \Phi \in S^{-}|q(\Phi) <0\}.
\end{equation}
\noindent
Both are acted on transitively by the real group 
Spin(6,6)$\times {\mathbb{R}}^{+}$. On those sets we define
the homogeneous degree two functional (also known as the
Hitchin functional)
\begin{equation}\label{Hitchinfunc}
\mathcal{H}(\Phi)  \overset{\text{def}}{=} 
\sqrt{\frac{-q(\Phi)}{3}}.
\end{equation}
\noindent
Note that if $\Phi = \varphi + \bar{\varphi}$ for a pure
spinor $\varphi$, 
\begin{equation}
i \mathcal{H}(\Phi) = < \varphi, \bar{\varphi}>.
\end{equation}
\noindent
The complex structure on $S^{\pm}_{q}$ is compatible with the 
Hitchin functional in the sense that 
\begin{equation}
i_{\Hat{\Phi}} < \cdot, \cdot > = d \mathcal{H}(\Phi),\;\;
\text{where}\;\; \Phi + i\Hat{\Phi} = 2\varphi.
\end{equation}
\noindent
Put into words, the Hamiltonian vector field of $\mathcal{H}(\Phi)$
is $\Hat{\Phi}$. Also, $ \Hat{\Hat{\Phi}} =- \Phi$. Now
$S^{\pm}_{q}$ are pseudo-K\"{a}hler spaces of signature (30, 2).
To make them K\"{a}hler, we define our moduli spaces as follows:
\begin{equation}
{\mathcal{M}}_{J^{1,2}} \overset{\text{def}}{=} S^{+}_q / 
{\mathbb{C}}^{*}
\cong \text{Spin}(6,6) \times {\mathbb{R}}^{+}/ SU(3,3) \times
{\mathbb{C}}^{*},
\end{equation}
\begin{equation}
{\mathcal{M}}_{{\Omega}^{1,2}}  \overset{\text{def}}{=} S^{-}_q 
/{\mathbb{C}}^{*}
\cong \text{Spin}(6,6) \times {\mathbb{R}}^{+}/ SU(3,3) \times
{\mathbb{C}}^{*}.
\end{equation}
\noindent
The local K\"{a}hler potentials are given by~\cite{Free}:
\begin{equation}
{\mathcal{K}}_{J^{1,2}} = - \ln \mathcal{H}({\Phi}^{+^{1,2}}) 
= -\ln(i<{\Phi}^{+^{1,2}}, {\bar{\Phi}}^{+^{1,2}}>),
\end{equation}
\begin{equation}
{\mathcal{K}}_{{\Omega}^{1,2}} = - \ln \mathcal{H}({\Phi}^{-^{1,2}}) 
= -\ln(i<{\Phi}^{-^{1,2}}, {\bar{\Phi}}^{-^{1,2}}>).
\end{equation}
\noindent
However, the resulting K\"{a}hler metric is not unique due to
the dilatino field. That ambiguity will be dealt with later.
For the time being we slightly modify the functionals by 
untwisting them so that we have ${\mathcal{H}}_0 : 
S^{\pm}_{q} \rightarrow {\Lambda}^6 V^*$. Then the volume
functional 
\begin{equation}
\mathbb{F} (\Phi) = \int_{Y}{\mathcal{H}}_0 (\Phi) 
\end{equation}
\noindent
has some remarkable properties. The critical points of
$\mathbb{F} (\Phi)$ are precisely the generalized Calabi-Yau
manifolds. Moreover, even though those critical points are 
never isolated (they constitute the orbits of the action of 
exact 2-forms), there are some that are degenerate only along
the orbits of the action of 2-forms, the so-called transversally
regular generalized Calabi-Yau manifolds. For those, the rank
of the Hessian is constant and always equal the difference 
between the dimension of the entire space and that of the 
orbit.  \\ 
\indent
Finally, we connect the $SU(3,3)$ spinors of Hitchin with the 
$SU(3) \times SU(3)$ structures~\cite{G-L-W}. Hereafter 
${\Phi}^{+^{1,2}} \in {\mathcal{M}}_{{\Omega}^{1,2}}$, while
${\Phi}^{-^{1,2}} \in {\mathcal{M}}_{J^{1,2}}$:
\begin{equation}
{\Phi}^{+^{1,2}}= {\eta}_{+}^{1,2}\otimes {\bar{\eta}}^{1,2}_{+}
\otimes e^{-\phi},
\end{equation}
\begin{equation}
{\Phi}^{-^{1,2}}={\eta}_{+}^{1,2} \otimes {\bar{\eta}}^{1,2}_{-} 
\otimes e^{-\phi}.
\end{equation}

\subsection{Mirror Moduli} 

The moduli spaces of pure Spin(6,6) spinors described, being
sweeping generalizations of Calabi-Yau moduli spaces, contain
many kinds of different elements. At one end of the spectrum
they incorporate classical Calabi-Yau manifolds, and allow for
mirror pairs thereof in ${\mathcal{M}}_{{\Omega}^{1,2}}$ and
${\mathcal{M}}_{J^{1,2}}$. Their matching is traditionally based
on the SYZ approach~\cite{SYZ}, and requires the existence
of $T^3$ fibrations. Swapping typical fibers with corresponding
dual ones effects mirror symmetry transformations~\cite{G-M-P-T}:
\begin{equation*}
e^{B +iJ^{1,2}} \longleftrightarrow {\Omega}^{1,2}.
\end{equation*}  
\noindent
Generalized Calabi-Yau manifolds generically do not possess any
$T^3$-fibrations, so that their pairs have to be matched via a 
different mechanism~\cite{G-M-P-T}. Indeed, Gra\~{n}a et al.
rely on supersymmetry transformations instead, to match mirror
pairs of generalized Calabi-Yau manifolds. There turn out to be
representation spaces (of closed differential forms) that are
swapped according to $\bf{8} + \bf{1} \leftrightarrow \bf{6}+\bf{3}$.
However, the bulk of ${\mathcal{M}}_{{\Omega}^{1,2}}$ and
${\mathcal{M}}_{J^{1,2}}$ consists of manifolds that are not
generalized Calabi-Yau, hence not supersymmetric at all.\\
\indent
We study those moduli with an eye towards symmetry (and
supersymmetry) properties of the ensuing 4-dimensional theories,
and might need to eliminate nonphysical variability inherent in
the compactifications. In the process we would greatly benefit
from identifying the transversal deformations of generalized
Calabi-Yau manifolds which form matching pairs in 
${\mathcal{M}}_{{\Omega}^{1,2}}$ and ${\mathcal{M}}_{J^{1,2}}$.
But quite apart from supersymmetry (and its breaking), the
question of establishing mirror symmetry between elements of
${\mathcal{M}}_{{\Omega}^{1,2}}$ and ${\mathcal{M}}_{J^{1,2}}$ is
of practical importance.\\
\indent
To start off, we get a grip on the moduli spaces viewing them
as unions of orbits of Spin(6,6).
Having  described the intrinsic torsion classes associated with
${\eta}^1, {\eta}^2$ in~\eqref{Itorsion1} -~\eqref{Itorsion2},
we delineate the set of all elements of Spin(6,6) preserving 
intrinsic torsion. 
For that, we look at the group extension $\aleph$, defined via 
\begin{equation*}
 ({\Lambda}^2 T^* Y)_{\textnormal{closed}} \longrightarrow
\aleph \longrightarrow {\textnormal{Diff}}^{\,e}(Y),
\end{equation*}
\noindent
where $\textnormal{Diff}(Y) \subset \textnormal{Diff}(M^{1,9})$ is a 
subgroup of the group of diffeomorphisms of the original manifold.
As such, $\aleph$  acts on the generic 6-dimensional manifolds $Y$
endowed with two $SU(3)$ spinors ${\eta}^1, {\eta}^2$ and leaves 
intrinsic torsion intact.
That action induces some linear actions on $TY$ and $T^*Y$, 
\textit{ergo} on
$TY \oplus T^*Y$. The latter action determines a representation of
$\aleph$ on $SO(6,6)$, which lifts to Spin(6,6). We denote its
image ${\aleph}_{\textnormal{linear}} \subset$ Spin(6,6). It is
a proper subgroup because $\mathfrak{so} (V \oplus V^*) =
V \otimes V^* \oplus {\Lambda}^2 V^* \oplus {\Lambda}^2 V$, and 
${\Lambda}^2 V$ does not preserve intrinsic torsion.
Furthermore, $V \otimes V^*$ exponentiates to a proper subgroup of
${\aleph}_{\textnormal{linear}}$ since there always are 1-forms that
are not closed.\\
\indent
The orbits of ${\aleph}_{\textnormal{linear}}$ constitute  
holomorphic foliations ${\mathcal{F}}_{J^{1,2}}$ and
${\mathcal{F}}_{{\Omega}^{1,2}}$. Using holomorphic Darboux
coordinates, we express the symplectic form as
\begin{equation}
<\cdot,\cdot>=\frac{i}{2}\sum_{l \in I} dz_l \wedge d{\bar{z}}_l.
\end{equation}
\noindent
Since the action of ${\aleph}_{\textnormal{linear}}$ is K\"{a}hler,
there are subsets of indices $I_0 (J) \subsetneq I$, $I_0 (\Omega)
\subsetneq I$ such that the respective symplectic forms on the 
leaves are
\begin{equation}
<\cdot,\cdot>|_{{\mathcal{F}}_{J^{1,2}}} =\frac{i}{2}\sum_{l \in I_0 (J)} 
dz_l \wedge d{\bar{z}}_l,
\end{equation}
\begin{equation}
<\cdot,\cdot>|_{{\mathcal{F}}_{{\Omega}^{1,2}}} =\frac{i}{2}
\sum_{l \in I_0 (\Omega)} dz_l \wedge d{\bar{z}}_l.
\end{equation}
\noindent
Therefore we define the Poisson structures as
\begin{equation}
{\pi}_{J^{1,2}} \overset{\textnormal{def}}{=} \frac{i}{2}
\sum_{l \in I_0 (J)}
{\partial}_l \wedge {\bar{\partial}}_l,\;\;\;
{\pi}_{{\Omega}^{1,2}} \overset{\textnormal{def}}{=} \frac{i}{2}
\sum_{l 
\in I_0 (\Omega)} {\partial}_l \wedge {\bar{\partial}}_l. 
\end{equation}
\indent
Now following Weinstein~\cite{We} we recall that a full dual
pair $(M_1, {\pi}_1) \overset{s_1}{\longleftarrow} M \overset{s_2}
{\longrightarrow}( M_2, {\pi}_2)$ consists of two Poisson 
manifolds $(M_1, {\pi}_1)$ and $(M_2, {\pi}_2)$, a symplectic
manifold $M$, and two submersions $s_1: M \rightarrow M_1$ and
$s_2: M \rightarrow M_2$ such that $s_1$ is Poisson, $s_2$ is 
anti-Poisson and the fibers of $s_1$ and $s_2$ are symplectic
orthogonal to each other. A Poisson (or anti-Poisson) mapping
is said to be complete if the pull-back of a complete
Hamiltonian flow under this mapping is complete. A full dual
pair is called complete if both $s_1$ and $s_2$ are complete.
The Poisson manifolds $M_1$ and $M_2$ are Morita equivalent if
there exists a complete full dual pair such that the 
submersions have connected and simply connected fibers~\cite{Xu}.
In the case of ${\mathcal{M}}_{J^{1,2}}$ and 
${\mathcal{M}}_{{\Omega}^{1,2}}$ we have the following full dual 
pair:
\begin{equation}\label{Morita}
({\mathcal{M}}_{J^{1,2}}, {\pi}_{J^{1,2}}) \longleftarrow
{\mathbb{R}}^{2n} \longrightarrow ({\mathcal{M}}_{{\Omega}^{1,2}},
{\pi}_{{\Omega}^{1,2}}),
\end{equation}
for $n$ large enough, and the intermediate space equipped with the 
standard symplectic structure. With those conditions satisfied,
\eqref{Morita} makes our moduli spaces Morita equivalent.\\
\indent
One property of Morita equivalent manifolds we utilize is that
their respective spaces of Casimir functions are isomorphic~\cite{We}.
That means the sheaves of Casimir functions are isomorphic too:
\begin{equation}
{\mathcal{C}}_{J^{1,2}} \cong {\mathcal{C}}_{{\Omega}^{1,2}}.
\end{equation}
\noindent
In the case of ${\mathcal{M}}_{J^{1,2}}$ and ${\mathcal{M}}
_{{\Omega}^{1,2}}$ these sheaves reduce to the holomorphic sheaves
of functions of the appropriate intrinsic torsion classes because
up to the actions of closed B-fields,
the classes provide complete sets of local first-order 
differential invariants of the structures involved~\cite{Sal}.
We denote them as
\begin{equation}
{\mathcal{W}}_{134} = \{f(W_1, W_3, W_4)\}^{\textnormal{hol}},\;\;\;
{\mathcal{W}}_{125} = \{f(W_1, W_2, W_5)\}^{\textnormal{hol}}.
\end{equation}
\noindent
As a consequence, ${\mathcal{W}}_{134} ={\mathcal{C}}_{J^{1,2}} 
\cong {\mathcal{C}}_{{\Omega}^{1,2}}={\mathcal{W}}_{125}$. By 
holomorphicity, their respective linear subsheaves are 
isomorpic as well, so that there is a correspondence
\begin{equation}
(W_1, W_3, W_4) \longleftrightarrow (W_1, W_2, W_5),
\end{equation}
\noindent
which essentially is a way of swapping $\bf{8} + \bf{1} 
\leftrightarrow \bf{6}+\bf{3}$.\\
\indent
Now Poisson vector fields are characterized as the fields
preserving the Poisson structure, i. e. for every Poisson $X$
we have ${\mathcal{L}}_{X} \pi =0$. The first Poisson 
cohomology groups are defined via
\begin{equation*}
H^1_{\pi}( \cdot) \overset{\textnormal{def}}{=} 
\frac{\textnormal{Poisson fields}}{\textnormal{Hamiltonian fields}}.
\end{equation*}
\noindent
Finally, applying the results of~\cite{G-G} we obtain
\begin{equation}\label{Poissonisomorphism}
H^1_{{\pi}_J}({\mathcal{M}}_{J^{1,2}})\cong {\check{H}}^1
({\mathcal{M}}_{J^{1,2}}, {\mathcal{W}}_{134}) \cong {\check{H}}^1
({\mathcal{M}}_{{\Omega}^{1,2}}, {\mathcal{W}}_{125}) \cong
H^1_{{\pi}_{\Omega}}({\mathcal{M}}_{{\Omega}^{1,2}}),
\end{equation}
\noindent
that is the first Poisson cohomology groups of our moduli
spaces are iso-/anti-iso-morphic to the respective first sheaf
cohomology groups, and the latter are canonically isomorpic.\\
\indent
Whenever there have been found classical matching Calabi-Yau 
elements $Y \in {\mathcal{M}}_{J^{1,2}}$, $\check{Y} \in
{\mathcal{M}}_{{\Omega}^{1,2}}$, we set the initial conditions for
the anti-isomorphic Poisson vector fields obtained from the Poisson
cohomology classes by projecting onto the $\pi$-transversal
subspaces (which is always possible because both ${\pi}_{J^{1,2}}$
and ${\pi}_{{\Omega}^{1,2}}$ are constant). Via mirror symmetry
their integral curves hitting $Y$ and $\check{Y}$ respectively
must match. One curious fact meriting further exploration is that
the numbers of integrable structures (i. e. of intersections of 
the Poisson integral curves with the leaves of ${\pi}_{J^{1,2}}$ and
${\pi}_{{\Omega}^{1,2}}$ that represent generalized Calabi-Yau
manifolds) in each moduli space must be equal.
By completeness, those numbers  
are precisely the numbers of such leaves, and the latter
are bounded by the Betti and Hodge numbers:
\begin{equation}
\# ({\mathcal{F}}_{J^{1,2}}|_{W=0}) \leqslant \sum_{i=1}^{b_Y}
\frac{b_Y !}{(b_Y - i)!},\;\;
b_Y = b^0(Y) +b^2(Y)+ b^4(Y),
\end{equation}
\begin{equation}
\# ({\mathcal{F}}_{{\Omega}^{1,2}}|_{W=0}) \leqslant \sum_{i=1}^{h_{\check{Y}}}
\frac{h_{\check{Y}} !}{(h_{\check{Y}} - i)!},\;\; \textnormal{where}
\end{equation}
\begin{equation}
h_{\check{Y}}= h^{1,0}(\check{Y}) + h^{3,0}(\check{Y}) + 
h^{2,1}(\check{Y}) + h^{3,2}(\check{Y}).
\end{equation}
\indent
Regardless of their enumeration, some leaves among the sets
${\mathcal{F}}_{J^{1,2}}|_{W=0}$ and ${\mathcal{F}}_{{\Omega}^{1,2}}|_{W=0}$
are transversally regular.\\
\indent
If no mirror pairs have been identified (and the above estimates
no longer apply), we can still use the
isomorphism~\eqref{Poissonisomorphism}. 
Namely, we can match thin stacks of symplectic leaves with
transversally regular generalized Calabi-Yau manifolds sandwiched in
the middle. Any two of those are
isomorphic. This effectively extends mirror symmetry to 6-manifolds 
that are gotten by transversal deformations of transversally
regular generalized Calabi-Yau ones.

\subsection{Nonspontaneous Supersymmetry Breaking}

So far we have been working with general Type II compactifications,
and the focus has been on their geometric properties. Now we turn
to a concrete Type II theory to see the interplay between those
geometric properties and supersymmetry. To fix ideas, we specialize
to Type IIA and IIB gauged supergravity and its democratic formulation
due to Bergshoeff at al.~\cite{B-K-O-R-VP}, presently grinding it down
further to exhibit a mechanism of supersymmetry breaking which does
not affect vanishing of the vacuum expectation values of 
gravitinos and other 4-dimensional fields. We are interested in the
situation where the effective theory has the minimal $N=2$
supersymmetry. In other words, we single out eight particular Type II
supersymmetries that descend to the effective theory. The
corresponding supersymmetry parameters $({\varrho}^1, {\varrho}^2)$
are gathered in the doublet denoted $\vec{\varrho}$. Then the most
general 10-dimensional gravitino supersymmetry transformation 
in the Einstein frame with all the R-R fluxes turned off has the form:
\begin{equation}
\delta {\Psi}_O =D_O \vec{\varrho} - \frac{1}{96} e^{-\frac{\phi}{2}}
({\Gamma}_O^{PQR}H_{PQR} -9{\Gamma}^{PQ}H_{OPQ})a^i {\sigma}_i
\vec{\varrho}.
\end{equation}
\noindent
Here $H_{OPQ}$ are the N-S flux coefficients, $O, P, Q \in \{0,..., 9\}$,
and $a^i$'s  are theory-specific. Because of the compactification
ansatz~\eqref{split} all fields on $M^{1,9}$ split. We write the 
ten-dimensional gamma matrices ${\Gamma}^M = 
({\Gamma}^{\mu}, {\Gamma}^m)$ as
\begin{equation}
{\Gamma}^{\mu} = {\gamma}^{\mu} \otimes {\bf{1}},\; \mu = \{0,1,2,3\},
\;\;\; {\Gamma}^m ={\gamma}_5 \otimes {\mathring{\gamma}}^m,\;
m=\{1, ..., 6\},
\end{equation}
\noindent
where ${\gamma}_5 = i{\gamma}^0{\gamma}^1{\gamma}^2{\gamma}^3$. 
In keeping with this splitting the ten-dimensional gravitinos
become ${\Psi}_M =({\Psi}_m, {\psi}_{\mu})$. To simplify matters, 
we set ${\eta}^1 ={\eta}^2 = \eta$, and decompose
the ten-dimensional supersymmetry parameters:
\begin{equation}
{\varrho}_{\textnormal{IIA}}^{1,2} = {\varrho}^{1,2}_{+} \otimes
{\eta}_{+} + {\varrho}^{1,2}_{-} \otimes {\eta}_{-},
\end{equation} 
\begin{equation}
{\varrho}_{\textnormal{IIB}}^{1,2} = {\varrho}^{1,2}_{+} \otimes
{\eta}_{-} + {\varrho}^{1,2}_{-} \otimes {\eta}_{+}.
\end{equation} 
\noindent
Abusing notation we let the `+' and `-' signs  signify both
four-dimensional and six-dimensional chiralities. Then 
supersymmetry imposes two gravitino and two dilatino equations: 
\begin{equation}
(D_m \pm \frac{1}{4}i_{{\partial}_m}H){\eta}_{\pm} =0,
\end{equation}
\begin{equation}
(D_m - d\phi \pm \frac{1}{2}H){\eta}_{\pm} =0.
\end{equation} 
\noindent
Those equations narrow down the list of supersymmetric
backgrounds. Thus $D_m {\eta}_{\pm} =0$ is a prerequisite. Hence
$H=0$, $\phi $ constant are the only choices. Therefore from
now on we fix our dilatino field to support on-shell supersymmetry.
Type IIA supersymmetric backgrounds are all $B$-transforms of symplectic
manifolds, while Type IIB ones are all complex~\cite{G-M-P-T}.\\
\indent
The supersymmetry transformations of ${\psi}_{\mu}$ are~\cite{G-L-W}:
\begin{equation}
\delta {\psi}^{1}_{\textnormal{IIA} \mu} 
=D_{\mu} {\varrho}^1  - ie^{\frac{1}{2}({\mathcal{K}}_J +
{\mathcal{K}}_{\Omega}) + \phi} <{\Phi}^{+}, d {\bar{\Phi}}^{-}>
{\varrho}^1,
\end{equation}
\begin{equation}
\delta {\psi}^{2}_{\textnormal{IIA} \mu} 
=D_{\mu} {\varrho}^2  + ie^{\frac{1}{2}({\mathcal{K}}_J +
{\mathcal{K}}_{\Omega}) + \phi} <{\Phi}^{+}, d {{\Phi}}^{-}>
{\varrho}^2,
\end{equation}
\begin{equation}
\delta {\psi}^{1}_{\textnormal{IIB} \mu} 
=D_{\mu} {\varrho}^1  - ie^{\frac{1}{2}({\mathcal{K}}_J +
{\mathcal{K}}_{\Omega}) + \phi} <{{\Phi}}^{-}, d {{\Phi}}^{+}>
{\varrho}^1,
\end{equation}
\begin{equation}
\delta {\psi}^{2}_{\textnormal{IIB} \mu} 
=D_{\mu} {\varrho}^2  + ie^{\frac{1}{2}({\mathcal{K}}_J +
{\mathcal{K}}_{\Omega}) + \phi} <{\Phi}^{-}, d {\bar{\Phi}}^{+}>
{\varrho}^2.
\end{equation}
\noindent
From these formulas we glean the expression for the mass
of gravitinos (maintaining the masses equal for ${\psi}^1$
and ${\psi}^2$):
\begin{equation}
m_{\textnormal{gravitino}} = c_{\textnormal{IIA, IIB}}
|\langle \text{VAC}|ie^{\frac{1}{2}({\mathcal{K}}_J +
{\mathcal{K}}_{\Omega}) + \phi} <{\Phi}^{\pm}, d {\Phi}^{\mp}>
|\text{VAC} \rangle|.
\end{equation}
\indent
Clearly, $d{\Phi}^{\pm} =0$ implies $m_{\textnormal{gravitino}} = 0$,
but there are other possibilities to get 
$<{\Phi}^{\pm}, d {\Phi}^{\mp}> =0$. We can always construct 
Lagrangian submanifolds (with respect to $<\cdot, \cdot>$) that
transversally intersect our foliations. 
Unfortunately, far away, there is no way of knowing whether 
the background manifold is supersymmetric, in
which case $m_{\textnormal{gravitino}} = 0$ trivially. But close
to transversally regular generalized Calabi-Yau manifolds all
transversally deformed Hitchin spinors are nonintegrable by 
transversal regularity.
Therefore plenty of background manifolds realize nonspontaneous
supersymmetry breaking. It must occur once the fluxes are in,
\textit{ergo} the moduli spaces are enlarged sufficiently.\\
\indent
To demonstrate pervasiveness of nonspontaneous supersymmetry 
breaking we now consider the space-time backgrounds that realize
only four supercharges. Then $N=2$ supermultiplets split. In
particular, the $N=2$ gravitational multiplet decomposes into an
$N=1$ gravitational multiplet containing the metric and one
gravitino $(g_{\mu \nu}, {\psi}_{\mu})$, and a $N=1$ spin-3/2
multiplet containing the second gravitino and the graviphoton
$({\psi}'_{\mu}, C_{\mu})$. The appearance of a standard $N=1$-
type action requires that the latter multiplet be projected out.
With all those conditions in place, the $N=1$ superpotentials
${\mathcal{P}}_{\textnormal{IIA}}, {\mathcal{P}}_{\textnormal{IIB}}$
have been computed from the supersymmetry transformation of the
linear combination of the two $N=2$ gravitinos which resides in 
the $N=1$ gravitational multiplet~\cite{G-L-W}. All such linear
combinations are parameterized by two angles that specify a
particular embedding of the $N=1$ submultiplet inside $N=2$.
They are
\begin{equation}
{\mathcal{P}}_{\textnormal{IIA}}=i{\cos}^2 \alpha e^{i \beta}
<{\Phi}^{+}, d{\bar{\Phi}}^{-}> - i{\sin}^2 \alpha e^{-i\beta}
<{\Phi}^{+}, d{\Phi}^{-}>,
\end{equation}
\begin{equation}
{\mathcal{P}}_{\textnormal{IIB}}= i{\cos}^2 \alpha e^{i \beta}
<{\Phi}^{+}, d{\Phi}^{-}> - i{\sin}^2 \alpha e^{-i\beta}
<{\Phi}^{+}, d{\bar{\Phi}}^{-}>.
\end{equation}
\noindent
Here transversal deformations furnish the space of $D$-flat 
directions thus eliminating spontaneous supersymmetry breaking.\\

\subsection{The $G$-Reduction}
In the previous subsection we have shown that sufficiently small
transversal deformations (within ${\mathcal{M}}_{{\Omega}}
\times {\mathcal{M}}_{{J}}$) of transversally regular generalized 
Calabi-Yau manifolds lead to nonspontaneously broken supersymmetry.
There still lingers the question whether the underlying mechanism
of supersymmetry breaking is that of Section 3.\\
\indent
To address that, we first observe that any effective 4-dimensional
theory is impervious to the actions of diffeomorphisms and closed
B-fields. Whence in the physical space-time $\eta$'s  transform
in accordance with the transversal deformations and are governed
by the (functions of) intrinsic torsion classes. Moreover, the
diagram below is commutative, and its vertical arrows indicate
surjective  mappings squashing the dilatino fields: 
\begin{equation*}
\begin{CD}
{\mathcal{M}}_{{\Omega}} \times {\mathcal{M}}_{{J}}
@>\text{Spin}(6,6)>> 
{\mathcal{M}}_{{\Omega}} \times {\mathcal{M}}_{{J}} \\
@VVV                 @VVV\\
{\Lambda}^*T^*Y \otimes \mathbb{C} @>\text{Spin}(6)>> 
{\Lambda}^*T^*Y \otimes \mathbb{C}
\end{CD}
\end{equation*}
\indent
Having made those points, we proceed to describe precisely how
$\eta$'s behave in the effective four-dimensional theory.
A convenient basis for all spinors on $Y$ is 
provided by~\cite{G-M-P-T}: $\{ \eta,\; {\mathring{\gamma}}\eta,\; 
{\mathring{\gamma}}^m \eta \}$. Relabeling the spinors we organize 
them into the following $SU(4)$-frame:
\begin{equation*}
\begin{bmatrix}
\eta + i{\mathring{\gamma}}\eta\\
\eta - i{\mathring{\gamma}}\eta\\
\theta + i{\mathring{\gamma}}\theta\\
\theta - i{\mathring{\gamma}}\theta
\end{bmatrix}.\;\;\;
{\textnormal{Now let}}\;\;\;
\begin{bmatrix}
\eta + i{\mathring{\gamma}}\eta\\
\eta - i{\mathring{\gamma}}\eta\\
\theta + i{\mathring{\gamma}}\theta\\
\theta - i{\mathring{\gamma}}\theta
\end{bmatrix}(t)
\end{equation*}
\noindent
be a one parameter family of spinor frames stemming from a
family of transversal deformations that realize nonspontaneous
supersymmetry breaking. Expanding in a Taylor series we get
\begin{equation}
\begin{bmatrix}
\eta + i{\mathring{\gamma}}\eta\\
\eta - i{\mathring{\gamma}}\eta\\
\theta + i{\mathring{\gamma}}\theta\\
\theta - i{\mathring{\gamma}}\theta
\end{bmatrix}(t)=
\begin{bmatrix}
\eta + i{\mathring{\gamma}}\eta\\
\eta - i{\mathring{\gamma}}\eta\\
\theta + i{\mathring{\gamma}}\theta\\
\theta - i{\mathring{\gamma}}\theta
\end{bmatrix}+
t \sum_{i} W_{il} M^l
\begin{bmatrix}
\eta + i{\mathring{\gamma}}\eta\\
\eta - i{\mathring{\gamma}}\eta\\
\theta + i{\mathring{\gamma}}\theta\\
\theta - i{\mathring{\gamma}}\theta
\end{bmatrix} + O(t^2),
\end{equation}
subject to the differential constraint
\begin{equation}d
\begin{bmatrix}
\eta + i{\mathring{\gamma}}\eta\\
\eta - i{\mathring{\gamma}}\eta\\
\theta + i{\mathring{\gamma}}\theta\\
\theta - i{\mathring{\gamma}}\theta
\end{bmatrix}=
\begin{bmatrix}
0\\
0\\
0\\
0
\end{bmatrix}.
\end{equation}
\noindent
Here $W_{il}$ are holomorphic functions of one of the 
intrinsic torsion classes, not all vanishing because of 
the transversality condition, and $M^l \in \mathfrak{su}(4)$
are nonsingular again by transversality.\\
\indent
Now by our extension of mirror symmetry~\eqref{Poissonisomorphism}, 
there must exist a diffeomorphism/closed B-field whose action relates 
$(W_2, W_5)$ with $(W_3, W_4)$ so that up to a scalar function
\begin{equation}
U(\sum_{i= \{1,2,5 \}}W_{il}M^l)U^H =\sum_{i= \{1,3,4 \}}W_{il}M^l,
\end{equation}
\noindent
which is an explicit realization of the representation swap
$\bf{6} + \bf{3} \leftrightarrow \bf{8} + \bf{1}$.
Using irreducibility we are forced to conclude that 
diffeomorphisms/closed B-fields cannot commute with $M^l$.
Therefore $M^l = \left[ \begin{smallmatrix}
0 & [\cdot ]\\
[\cdot ]& 0 \end{smallmatrix} \right]$, and up to the action of 
diffeomophisms/closed B-fields $M^l =K_l$. And the 
smallest subgroup of $SU(4)$ incorporating $K_l$ is $G$. Thus
the connection matrices are determined and the connection
~\eqref{nabla} takes the form 
\begin{equation}
P_{\mu} \longrightarrow i{\nabla}_{\mu}(x) 
\overset{\text{def}}{=} i({\varepsilon}^{\nu}_{\mu}(x)
{\partial}_{\nu} + i\sum_{j} W_{j\mu}^a (x) K_a),
\end{equation}
\noindent
where the potentials $W_{j\mu}^a (x)$ are defined via~\eqref{split}
and~\eqref{fibration}. For instance, within a supersymmetric
region $W_{j\mu}^a (x) \equiv 0$, hence ${\varepsilon}^{\nu}_{\mu}(x)
\equiv {\delta}^{\nu}_{\mu}$.

\end{document}